\documentstyle[aps,epsf,preprint]{revtex}    
\tightenlines
\begin{document}

\title{Apparent suppression of turbulent magnetic dynamo action by a dc
magnetic field} 

\author{David C. Montgomery}

\address{
         Department of Physics and Astronomy,
         Dartmouth College, Hanover, New Hampshire 03755-3528, USA.}
\author{
         W. H. Matthaeus,
         L. J. Milano and
         P. Dmitruk  }

\address{
         Bartol Research Institute, University of Delaware,
         Newark, Delaware 19716, USA.}

\maketitle

\begin{abstract} 

Numerical studies of the effect of a dc magnetic field on dynamo
action (development of magnetic fields with large spatial scales),
due to helically-driven magnetohydrodynamic turbulence, are reported.
The apparent effect of the dc magnetic field is to suppress the
dynamo action, above a relatively low threshold. However,  
the possibility that the suppression results from an improper
combination of rectangular triply spatially-periodic boundary
conditions and a uniform dc magnetic field is addressed: heretofore a common and
convenient computational convention in turbulence investigations.
Physical reasons for the observed suppression are suggested.
Other geometries and boundary conditions are offered for which the
dynamo action is expected not to be suppressed by the presence of
a dc magnetic field component.

\end{abstract}

\newpage
\section{Introduction} 

	The spontaneous development of large-scale magnetic fields in
 astrophysical or geophysical settings has been a problem of interest
 at least since the time of Gauss, and remains imperfectly understood.
 One promising candidate for a likely basic and universal ``dynamo''
 process has been the turbulent inverse cascade of magnetic helicity
 \cite{FrischEA75,PouquetEA76,Pouquet87}, a topic which
 has received considerable attention among turbulence theorists. In
 broad terms, a mechanical source of helical turbulent excitations
 (often combining thermal convection and rotational properties of the
 system) is conjectured to excite magnetic turbulence at small spatial
 scales by a variety of mechanisms such as flux tube stretching. The
 magnetic turbulence then inversely cascades toward ever-larger scales
 because of certain statistical properties of the magnetohydrodynamic
 (MHD) equations, properties that are mathematically understandable
 but are not altogether intuitive. Virtually no experimental
 verification of the mechanism has been available, and most of the
 evidence for the process that has accumulated has come largely from
 computationally-intensive numerical solutions of the MHD equations,
 most notably those of Meneguzzi {\it et al.} \cite{MeneguzziEA81}. 
 The effect is
 inherently three-dimensional (3D), because of the involvement of magnetic
 helicity, although an analogous process driven by
 inverse transfer of mean square magnetic potential (mean square flux function)
 is observed
 in two-dimensional (2D) computations and simulations.

	The situation considered is related to, but not quantitatively well
 represented by, the so-called ``alpha effect'',  which necessarily
 assumes large gaps in the magnetic energy spectrum across which the
 magnetic excitations are supposed to jump as a consequence of
 microscopic nonlinear processes \cite{Moffatt78,KrauseRaedler80,FieldEA99}.
 Such numerical evidence as has
 been presented has always shown that any such initial spectral gap
 quickly fills in and disappears, and the energy transfer into any
 band in wavenumber space thereafter tends to be from adjacent
 wavenumber bands, not from remote ones. Nevertheless, the alpha
 effect should be considered to stand as the precursor or first hint
 of dynamo action through the inverse cascade of magnetic helicity.

	The investigations reported here were intended to explore the effect
 of an imposed dc magnetic field on the dynamo action associated with
 the inverse magnetic helicity cascade. Sometimes in astrophysical
 cases, and particularly in laboratory situations such as the
 reversed-field pinch, MHD turbulence takes place in the presence of
 strong dc magnetic fields whose source currents are external to the
 magnetofluid. Numerical verification of inverse helicity cascades and
 their attendant dynamo action of the kind reported by Meneguzzi {\it et al.}
 \cite{MeneguzziEA81} have generally been achieved by computations in
 three-dimensional rectangular periodic boundary conditions involving
 no such imposed dc magnetic fields. It is of interest to see how the
 inverse helicity cascade and the
 supposed dynamo process might be affected by the presence of
 externally-imposed dc magnetic fields, particularly since it is known
 that dc magnetic fields strongly affect MHD turbulence in other
 contexts, rendering it highly anisotropic 
 \cite{ShebalinEA83,OughtonEA94,MontgomeryMatthaeus95,MilanoEA01}.
 Some recent computations \cite{CattaneoHughes96}
 examine effects of an external field on
 the alpha parameter, but not its effects on inverse cascade.
 The subject appears to be far from complete, and considerable controversy
 persists regarding the nature of the dynamo saturation process
 \cite{CattaneoHughes96,BlackmanField00,BrandenburgDonner97}.
 For this reason, we have undertaken
 the apparently straightforward task of repeating the computations of
 Meneguzzi {\it et al.} \cite{MeneguzziEA81} but in the presence of an 
 imposed spatially
 uniform dc magnetic field, to see what changes that field would
 produce.

	In Sec. II, we report the results of this computation. At first
 sight, they are surprising in that the presence of the dc magnetic
 field effectively suppresses the dynamo action at fairly low levels,
 in this conventional
 framework that has become standard for MHD turbulence theory and
 computation. 

 In Sec. III we discuss the
 reasons why we believe the conclusions are dependent on certain
 artificialities and inconsistencies imbedded in the standard
 formulation's geometry in this case. Finally, in Sec. IV, we offer
 suggestions for future possibilities for turbulence computations in
 alternative geometrical settings where we believe a more accurate set
 of conclusions concerning the effects of dc magnetic fields in the
 dynamo problem may be drawn.

\section{APPARENT DYNAMO SUPPRESSION} 

 We employ a fully dealiased spectral code
 \cite{GottliebOrszag77}
 for solving the equations of incompressible MHD with uniform
 mass density. These are basically an equation of motion, Faraday's
 law, and an Ohm's law, with velocity fields and magnetic fields which
 are both solenoidal. Written out in detail, we have, 

\begin{equation}
   \partial_t {\bf v} + {\bf v}\cdot\nabla {\bf v} =
   -\nabla p +  {\bf j} \times {\bf B} + \nu \nabla^2 {\bf v} + {\bf f}
   \label{navier-stokes}
\end{equation}
\begin{equation}
   \partial_t {\bf B} =
   - \nabla \times {\bf E}
   \label{faraday}
\end{equation}
\begin{equation}
   {\bf E} + {\bf v} \times {\bf B} = \eta {\bf j}
   \label{ohm}
\end{equation}
 The symbols are as follows. ${\bf B}$  is the magnetic field
 (in Alfv\'enic velocity units),
 ${\bf v}$ is the
 velocity field, and p is the pressure. ${\bf E}$ is the electric field
 and {\bf j}
 is the electric current density, given by the curl of ${\bf B}$. The
 dimensionless viscosity and resistivity are $\nu$ and $\eta$,
 respectively and are in effect the reciprocals of Reynolds-like
 numbers, mechanical and magnetic. For the present computations,
 $\nu$ and $\eta$
 are being chosen equal (unit magnetic Prandtl number). Both
 ${\bf B}$  and ${\bf v}$ have zero divergences. ${\bf B}$
 could be written as the curl of a
 vector potential ${\bf A}$, if desired. We will not greatly emphasize the
 role of ${\bf A}$, because of ambiguities in its role in possible definitions
 of magnetic helicity in the presence of a dc magnetic field
 \cite{Berger97,StriblingMatthaeus94}. These
 ambiguities are not central to the points we intend to make, which
 pertain primarily to the appearance or non-appearance of long-wavelength
 magnetic field spectral components -- directly computable without
 reference to ${\bf A}$. The inhomogeneous mechanical forcing term ${\bf f}$,
 on the right hand side of Eq. (1), is a given, solenoidal, random,
 vector function designed to inject mechanical helicity and is
 intended to mimic whatever mechanically turbulent processes one may
 invoke as the source of the velocity-field excitations.


 We attempt to
 work in regimes in which the mean kinetic energy per unit volume and
 rms velocity field are of order unity, so that our Reynolds-like
 numbers are based on length scales that are roughly one sixth of the
 basic box size.
 All fields are expanded in three-dimensional Fourier series in a
 cubical box of edge length $2\pi$ , so that rectangular periodic
 boundary conditions are being assumed in all three directions. The
 magnetic field will be written as ${\bf B} = {\bf B_0} +  {\bf b}$, where
 ${\bf B_0}$ is a uniform
 constant and the spatial average of the variable magnetic field, {\bf b},
 is zero.

	It has often been thought desirable to study MHD turbulent processes at
 as high Reynolds-like numbers as possible (i.e., $\nu$ and $\eta$ as
 small as possible), but in three dimensions, long-time computations
 of this type require high spatial resolution and make expensive
 demands on computer time. We have concluded that the points we wish
 to make here can be made convincingly at lower Reynolds-like numbers,
 and the runs to be reported have all been carried out with $\nu$ and $\eta$
 $= 0.01$, while spatial resolution of only 32-cubed has been
 employed.  
 It would
 be desirable to repeat these runs with lower $\nu$ and $\eta$ and with
 greater spatial resolution, but we would not expect the conclusions
 we will draw here to be changed by doing so.

	We will illustrate the results of many similar runs by showing
 omni-directional magnetic energy spectra for two identically-driven
 runs with $B_0 = 0$ (Fig. 1) and $B_0 = 1.0$ (Fig. 2), respectively.
 In contrast to previous studies that have focused almost exclusively on
 the alpha parameter \cite{CattaneoHughes96}, our emphasis will be
 on examination of the behavior of spectral quantities in response to
 dynamo action and the dc magnetic field strength.
 In particular the inverse cascade, if it occurs,
 involves transfer of magnetic energy to wavelengths longer than
 the mechanical forcing scales. The
 forcing function ${\bf f}$ can be adjusted to inject mechanical
 excitations whose helicity varies between zero and the maximum
 possible for the chosen mean forcing amplitude.
 [To accomplish this,
 we select each Fourier mode ${\bf v}({\bf k})$ with $\bf k$ 
 in the forcing band, and
 decompose into positive and negative mechanical helicity
 contributions (Chandrasekhar-Kendall functions).
  Each real degree of freedom is given a random increment scaled to
  $F \Delta t (1+\sigma)/2 $ for the positive helicity amplitude,
  or to  $F \Delta t (1-\sigma)/2$ for negative helicity modes.
  Forcing strength is controlled by $F$ (typically $1$),
 while the helicity injection is regulated by $\sigma$. 
 The timestep is $\Delta t$.
 For the runs shown in Figs. 1 and 2, $\sigma = 0.8$ corresponding to roughly
 $80$\% of the injected energy going into positive helicity modes.
 The forcing band in these runs
 includes all wave numbers between 5.0 and 5.3 in magnitude; a total of 67
 independent complex vector amplitudes are driven. At the
 beginning, the energy spectra are both empty.

 In Figs. 1 and 2 the solid lines are
 the mechanical energy spectra and the broken lines are the magnetic
 energy spectra, both at times $t=1000$.
  The unit of time is one eddy turnover time, based on unit
 length scale and unit rms velocity. The forcing amplitude $F=1$ is tuned so
 that the rms velocity field will be of order unity at the end of the
 run (saturation). The peaks in the spectra occur
 at the forcing band, where the mechanical excitations are being
 injected. At high
 wavenumber the magnetic energy exceeds the kinetic energy by a small amount, as is typical
 of MHD turbulence. Near the forcing wavenumber,
 kinetic energy dominates, reflecting the nature of the forcing.
 At wavenumbers lower than the forcing band, the spectra differ greatly for the
 two runs shown.

	Fig. 1, having $B_0 = 0$, essentially reproduces the results of
 Meneguzzi {\it et al.} \cite{MeneguzziEA81}. 
 It will be seen that the longest wavelength
 magnetic energy components have grown to more than an order of
 magnitude greater, in energy, than the velocity-field components that
 are driving them. The longest wavelength allowed by the boundary
 conditions dominates the spectrum. Everything below the forcing band
 in k-space represents spectral back-transfer to longer wavelengths.
 Fig. 2, with $B_0=1$ on the other hand, shows no such accumulation of energy of
 either kind at the longest wavelengths, which appear to be dominated
 by low-amplitude Alfv\'en waves that imply near-equipartition of energy
 between the magnetic and kinetic spectra.  The dynamo action at long
 wavelengths has been suppressed. Saturation has been achieved for both
 cases well before the time depicted.

	Figs. 3 and 4 show time histories of
 kinetic and magnetic energies for the same two runs in Figs. 1 and 2.
 For $B_0= 0$, Fig. 3 shows that magnetic energy overtakes kinetic energy 
 at about
 $t=75$ and thereafter the system as a whole is magnetically dominated.
 This is due to the strongly enhanced magnetic energy at
 the longest allowed wavelength as seen in Fig. 1.
 For $B_0=1$, Fig. 4 shows that magnetic energy saturates at a lower level,
 about 25 - 30 \% less than the kinetic energy. In this ``equipartition''
 regime, referring to Fig. 2, there is clearly no buildup of longest wavelength
 magnetic energy in the largest scale modes.
 Finally we note that buildup of the magnetic energy, or the absence thereof,
 may be associated with the generation and long wavelength buildup of magnetic
 helicity, or its absence. This is evidenced by Fig. 5, showing
 normalized magnetic helicity and kinetic
 helicity spectra at $t=1000$ for $B_0 = 0$, and by Fig. 6, showing the same 
 quantities
 $B_0 = 1$. For $B_0=0$ this conclusion is completely consistent with
 Meneguzzi {\it et al.} \cite{MeneguzziEA81}, except that we see it as a 
 saturated effect and at much later times.
 Other runs (details not shown) have been done with non-helical ($\sigma=0$)
 mechanical forcing, and dynamo action
 such as that seen in Figs. 1 and 5 is not observed.
 We have defined magnetic helicity here as the volume integral of 
 ${\bf a}\cdot{\bf b}$, where the ${\bf b}$ does not contain ${\bf B}_0$, 
 and is given by ${\bf b} = \nabla\times {\bf a}$. The mechanical helicity is 
 defined as the volume integral of the dot product of velocity and 
 vorticity, in the usual way.

	In still other runs, we have explored the effect of lowering $B_0$ in
 the presence of a fixed helical random forcing to see when and if the
 dynamo action would reappear. We did find a threshold value,
 somewhere between $B_0 = 0.1$ and $0.03$ in the dimensionless units, in
 which long-wavelength magnetic helicity-driven dynamo action reappeared.
 Below this value, the
 system evolved substantially as in Figs. 1 and 3, with a slight
 temporal offset. We do not hazard a guess as to what physical
 parameters the threshold may be dependent upon, because there are too
 many: $\nu$, $\eta$, $B_0$, the intensity of forcing, the location of
 the forcing band in k-space, and so on.

\section{ A RECONSIDERATION OF BOUNDARY CONDITIONS } 

	There had previously been some unexplained but not widely noticed
 features of turbulent MHD behavior in 3D rectangular periodic
 boundary conditions that suggested less than a complete understanding
 of the role the geometry was playing. In an alpha effect calculation
 generalized to the case of a uniform dc magnetic field, Montgomery
 and Chen \cite{MontgomeryChen84} had found an amplification matrix 
 whose trace tended
 to zero when $B_0$ became large, indicating  a less
 and less efficient alpha amplification for larger and larger imposed
 dc magnetic field. A 3D periodic computation of decaying MHD
 turbulence in a dc magnetic field 
 \cite{StriblingMatthaeus94} 
 showed no
 tendency for long wavelength helical magnetic field components to
 persist in the fashion they would when $B_0$ was zero. Both behaviors
 are consistent with, though they do not imply, the behavior reported
 in Sec. II for the driven case.

	Reluctantly, we have come to conclude that certain features of the
 combination of a dc magnetic field and rectangular periodic boundary
 conditions are unsatisfactory, and that these (computationally very
 convenient) boundary conditions, rather than the inherent physics,
 are controlling the computed behavior (see \cite{MontgomeryBates99}).
 In Ref.\cite{MontgomeryBates99},  reasons for distrusting triply-periodic 
 boundary conditions as adequate for this problem were spelled out 
 in considerable detail, and it has seemed unnecessary to reproduce them here.
 What seems to occur when the dc magnetic field is present and
 helical driving occurs is that the excitations built up are
 essentially Alfv\'en waves of a preferred helicity. A net emf builds up
 due to their attempts at dynamo action, corresponding to a non-zero
 spatially-averaged electric field parallel to ${\bf B_0}$. In nature, this
 would seek to drive a net current along ${\bf B_0}$, creating more magnetic
 flux in the perpendicular directions, but the rectangular periodic
 boundary conditions combined with Ampere's law permit no net current
 to flow through the system.  ``Open circuit'' boundary conditions have
 been effectively imposed. In a physical plasma, what would result
 from this would be a migration of electrons to one face of the box
 (normal to ${\bf B_0}$) and a net positive charge would appear on the opposite
 face, screening out the mean interior electric field parallel to ${\bf B_0}$.
 But that, too, is forbidden by the rectangular periodic boundary
 conditions.

 Clearly there are limitations on the physics that can be
 represented by periodic boundary conditions, and care must be employed
 in using and interpreting them 
 \cite{StriblingMatthaeus94,MontgomeryBates99,CattaneoHughes96,BlackmanField00}.
 It is necessary to recall that in electromagnetic theory,
 the theorems are for finite-sized systems whose fields fall off at
 infinity.  ``Infinite'' systems are a convenient idealization, 
 when considering, e.g.,  long current-carrying wires or parallel-plate
 capacitors, but it is still necessary to be able at least to imagine
 the infinite system as a limit of a finite one which becomes large.
 Similarly, the approximation of spatial homogeneity, often associated with the
 periodic model, is at best a local approximation. Moreover, attempts to
 extend homogeneity into the infinite domain limit may be fraught with
 difficulties, especially for MHD.
 The need for source currents for ${\bf B_0}$ somewhere, for each box in a
 triply-periodic array, clouds the picture of just what system it is
 that could be idealized by a periodic box that repeats itself
 indefinitely in all directions with a uniform dc magnetic field
 simultaneously present.
 Consequently a model consisting exclusively of periodic cells is too
 idealized to represent the entirety of a physical system, especially one like a dynamo
 in which small scales must communicate dynamically with very large scales
 There appear to be two possible solutions to this difficulty, 
 which are related.
 First, the periodic system might be embedded in a larger system, using some version of
multiple scale analysis to connect them \cite{StriblingMatthaeus94,Matthaeus99}.
In such an approach, periodic physics is a local effect and large 
scale dynamical
processes such as dynamo action will be taking place in the ``outer'' model,
itself not periodic. This approach, unless completed, cannot answer the question of whether
 dynamo action occurs, since the currents generated by the
 local $\langle {\bf v} \times {\bf b}\rangle$ lie outside the periodic model.
 Unless the full model is solved we do not know if such current could support 
 amplification of
 the large scale field and at what level.

  A more complete approach would be to formulate
 the dynamo problem
 in its entirely at the onset, in a framework that is not periodic.
 In the next Section, we consider whether it
 may not be possible, by incorporating more realistic boundary
 conditions, to do dynamo computations that do not experience what
 seems to be an artificial suppression in the presence of a mean dc
 magnetic field.

\section{ DISCUSSION AND FUTURE POSSIBILITIES } 

	We should stress that our goal here is physical understanding of one
 likely dynamo process for attaining long-wavelength magnetic fields
 out of turbulent microscopic MHD processes. For that reason, we have
 not addressed ourselves to the many admirable efforts at specific
 simulations of solar or geophysical magnetic fields, which
 necessarily incorporate many processes and effects omitted here, and
 whose success or failure is generally judged by the simulations'
 capacity for reproducing a wide range of observational features, such
 as sunspot patterns, solar prominences, periodicities, etc. We are
 focusing instead on only one MHD turbulent process in isolation, in
 an effort to understand it more correctly.

	If triply periodic rectangular boundary conditions are to be given up,
 it is natural to ask what the other possible geometries and boundary
 conditions there are for asking fundamental questions about MHD
 dynamos.  A first answer might be, spheres or disks
 \cite{BrandenburgDonner97}
 for astrophysical
 situations, and toroids or periodic cylinders [21-24] for laboratory ones. In
 all three cases, there is at least one coordinate, the radial one,
 which cannot be periodic. Turbulent inverse cascades of
 magnetic excitations in response to small-scale mechanical stimuli
 are readily imaginable in all three cases, though some symmetries
 that are convenient, such as isotropy and homogeneity, will be lost.
 Some experience has already been accumulated with weak MHD turbulence
 (though not for dynamo problems) in a rigid-wall, straight-cylinder
 geometry \cite{ShanEA91,ShanMontgomery93a,ShanMontgomery93b,ShanMontgomery94}.
 Tractable boundary conditions for finite magnetofluids are
 non-trivial and need to be debated, but incorporating them seems a
 likely direction in which to proceed to try and formulate the proper
 problem. The three cases above, in the incompressible limit, all have
 natural expansion bases for Galerkin-method computations
 (Chandrasekhar-Kendall eigenfunctions of the curl) that suggest
 themselves and have been used to some effect in the past.

	We conjecture that in more realistic geometries, finite in at least
 two dimensions, it will turn out that the presence of dc magnetic
 fields whose sources do not necessarily lie inside the magnetofluid
 do not act as a barrier to dynamo action of the inverse cascade type.
 That is, we do not expect long wavelength dynamo action in such
 geometries to be shut down by the presence of externally-supported dc
 magnetic fields. Only a formidable computational effort stands
 between us and a detailed answer.
 We re-emphasize that what we are offering is a conjecture, and not a fact; 
 and we stress again the need for driven turbulent MHD computations inside 
 finite geometries with imposed dc magnetic to reinforce or refute the 
 conjecture.

\acknowledgements 

This research supported in part by NSF grant
ATM-9713595.

\newpage

\begin{figure}
\caption[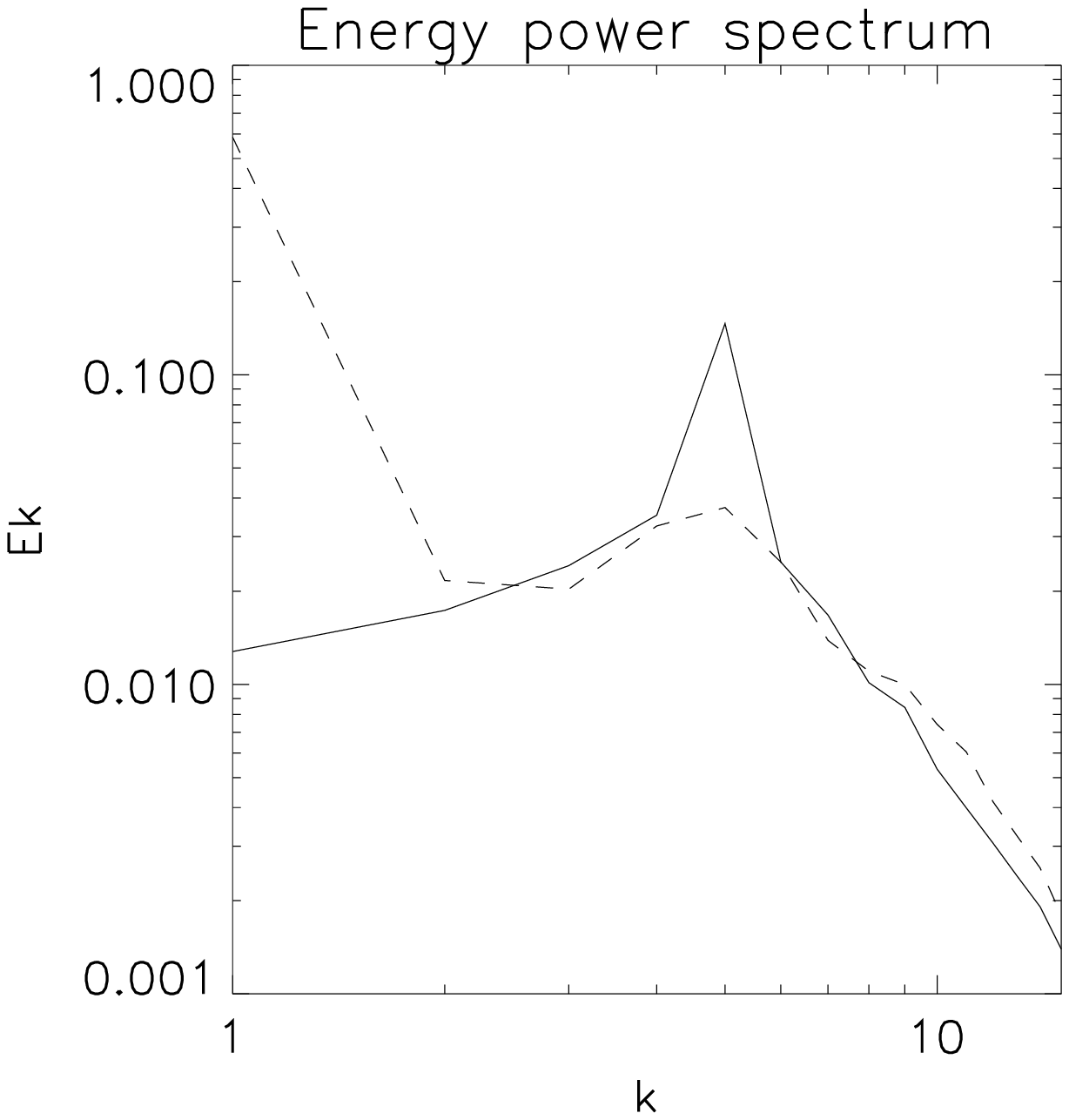]{
 Evolved ($t=1000$) kinetic energy spectrum (solid line) and magnetic
 energy spectrum (dashed line) in the presence of no dc magnetic field
 ($B_0 = 0$) with helical mechanical forcing.
 This spectrum is totally dominated by the $k=1$ modes.
\label{fig_1}}
\end{figure}

\begin{figure}
\caption[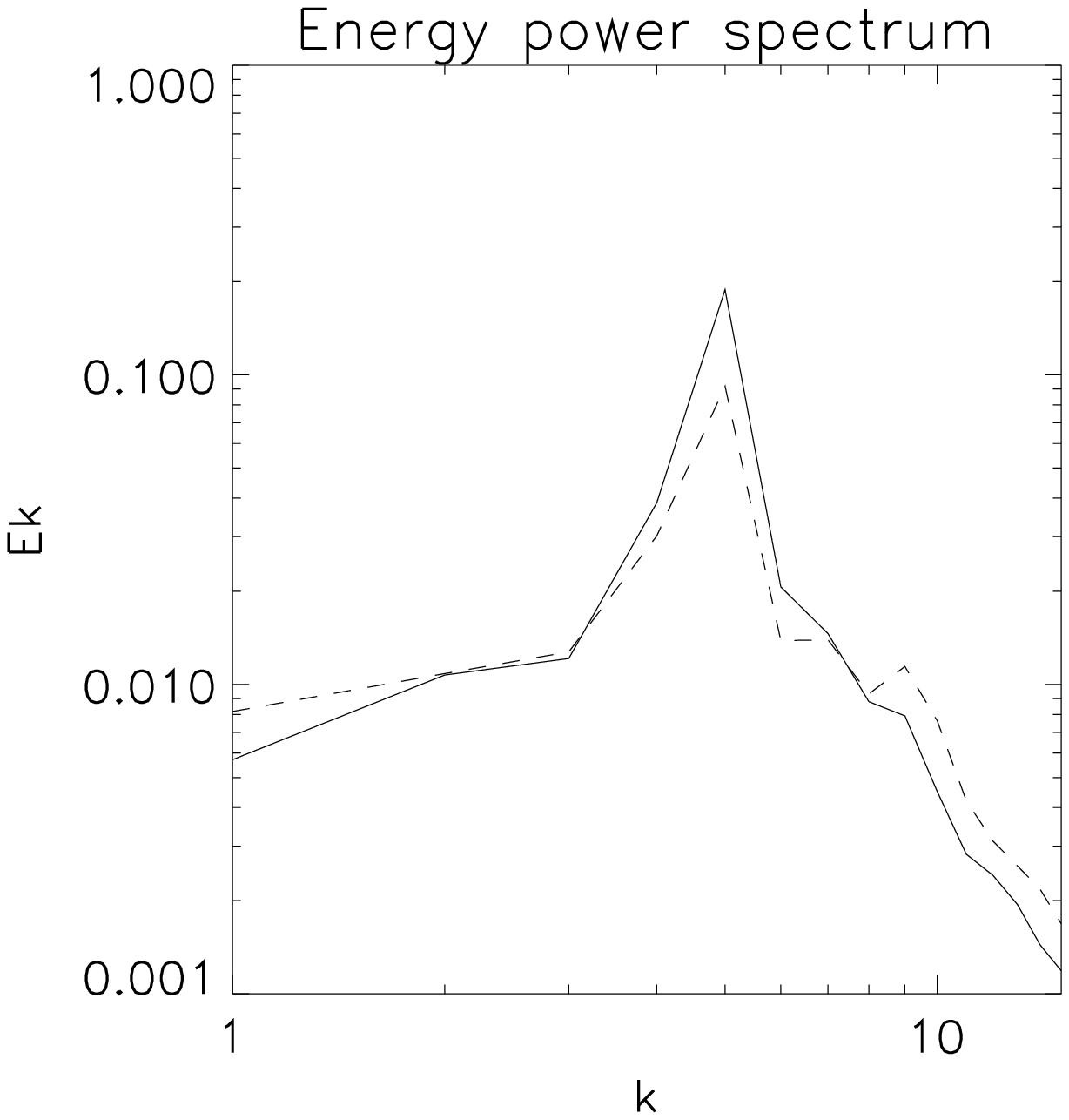]{
 Evolved ($t=1000$) kinetic energy spectrum (solid line) and magnetic
 energy spectrum (dashed line) in the presence of a dc magnetic field
 strength $B_0 = 1.0$ with helical mechanical forcing.
\label{fig_2}}
\end{figure}

\begin{figure}
\caption[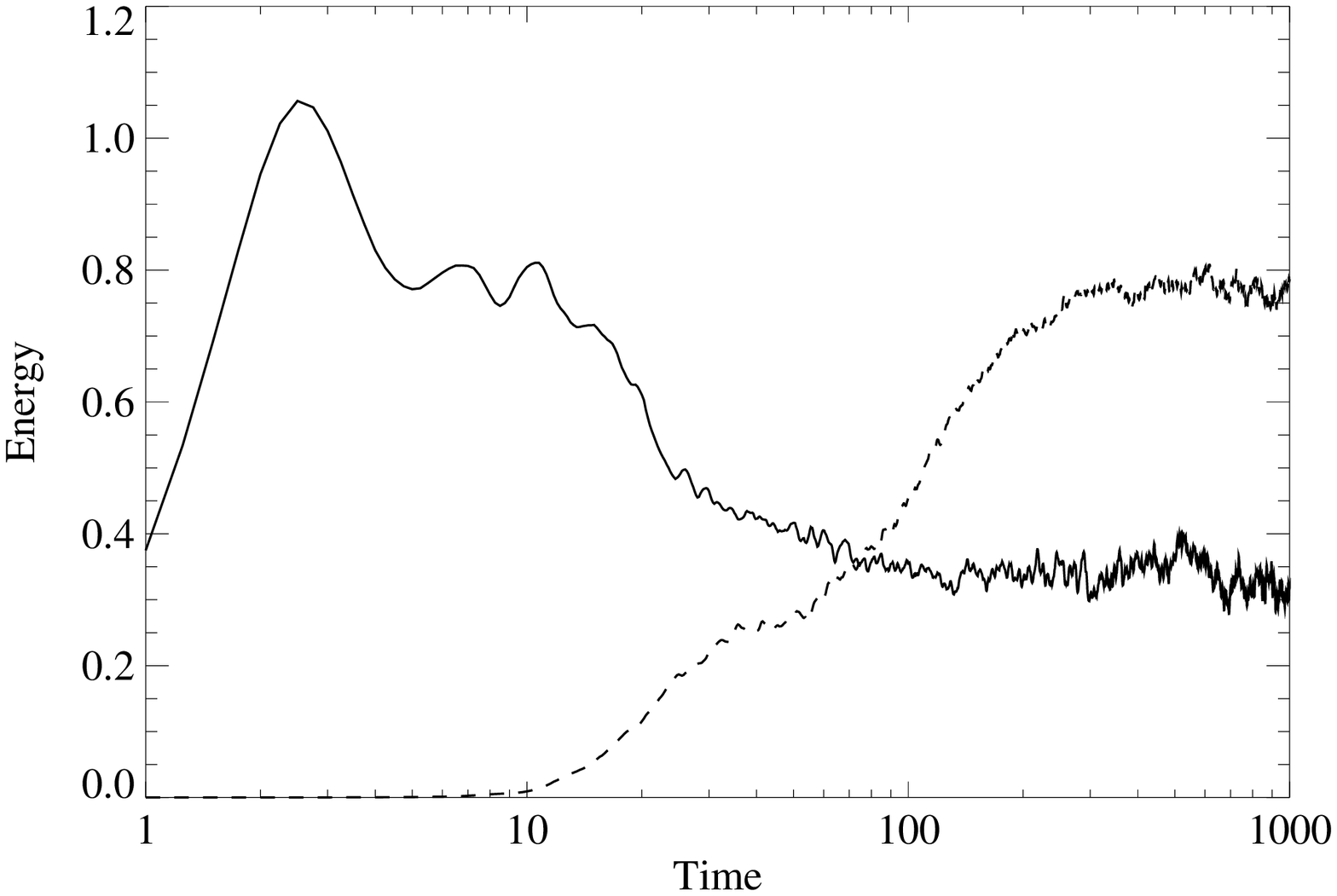]{
 Time histories of kinetic (solid line) and magnetic energy (dashed line)
        for the zero dc field case.
 \label{fig_3}}
\end{figure}

\begin{figure}
\caption[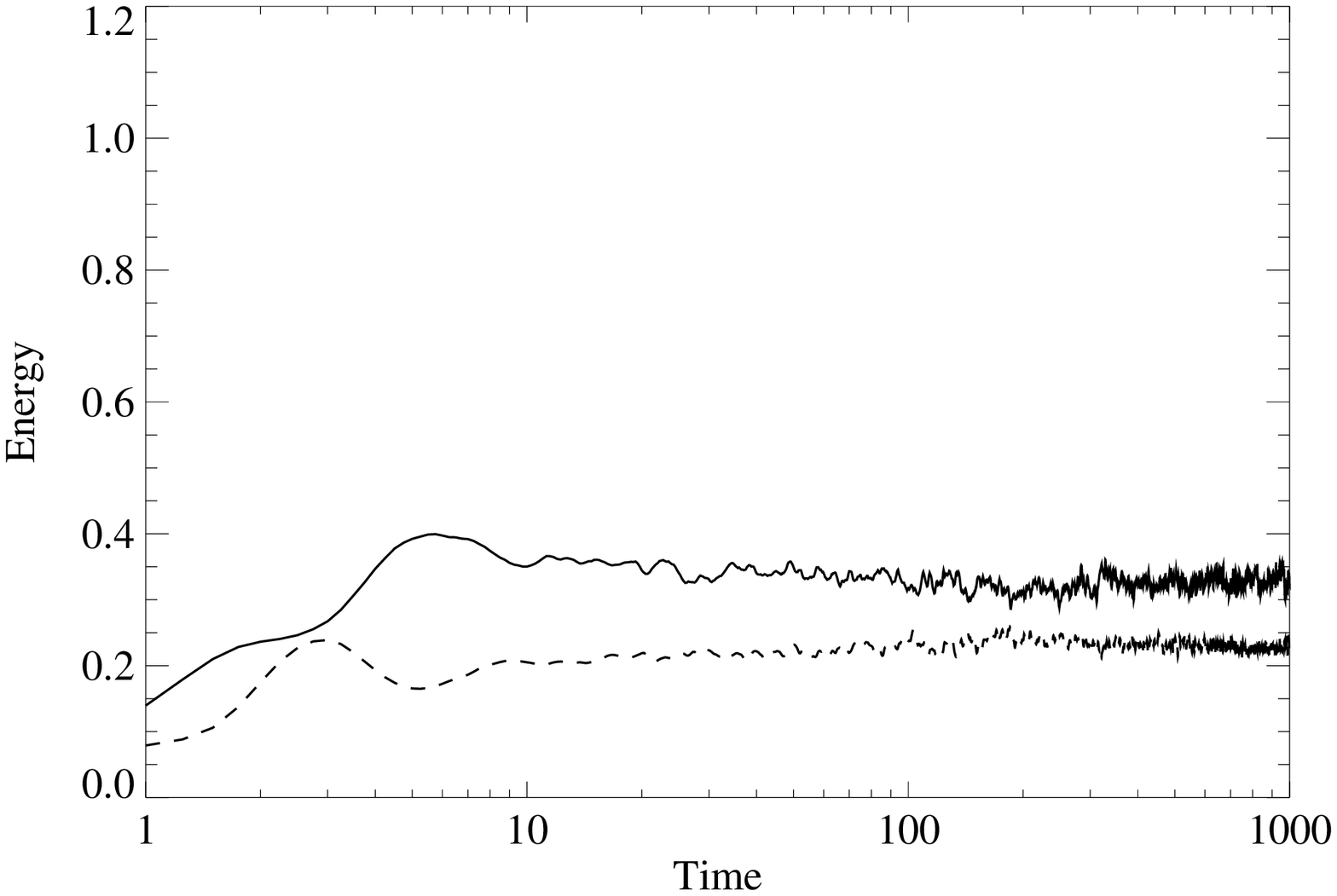]{
 Time histories of kinetic (solid line) and magnetic energy (dashed line)
         for the strong dc field case.
\label{fig_4}}
\end{figure}

\begin{figure}
\caption[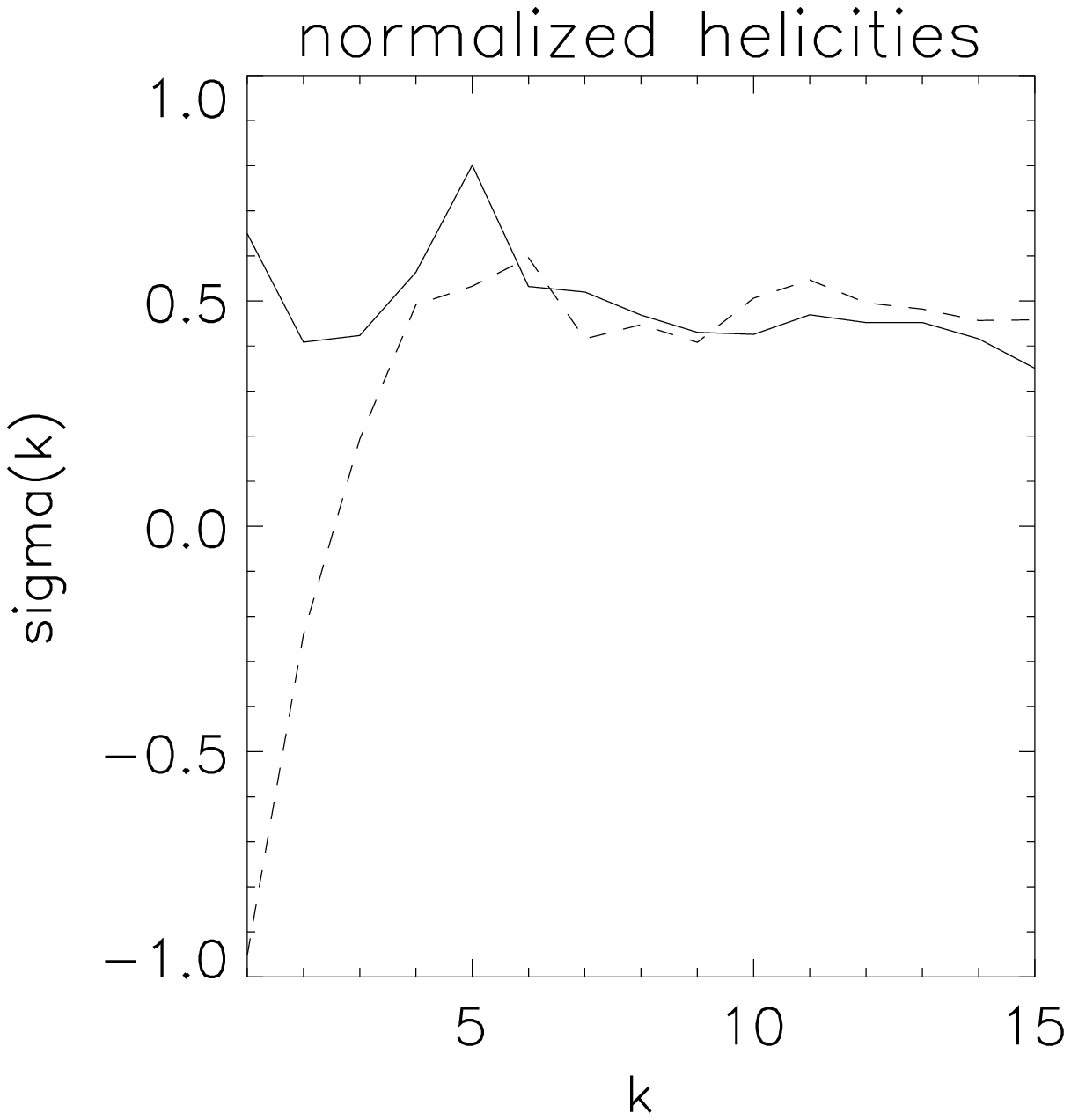]{
 Highly helical magnetic structure at the longest wavelength with $B_0=0$. 
 Normalized
 kinetic and magnetic helicities for the zero dc magnetic field case at $t=1000$.
In each case the associated gauge invariant normalized helicity is
$(E^+-E^-)/(E^++E^-)$ where $E^+$ and $E^-$ are the
decomposition of the respective energy into positive and negative
helicity contributions in the relevant wavenumber range.
Normalized helicities are nearly equipartitioned in forcing band
 and high wavenumber regimes. Note the very strong {\it negative} magnetic 
 helicity at the longest wavelength.
Combined with the magnetic energy spectrum in Fig. 1, this indicates a strong inverse cascade driven by
 magnetic helicity.
 \label{fig_5}}
\end{figure}

\begin{figure}
\caption[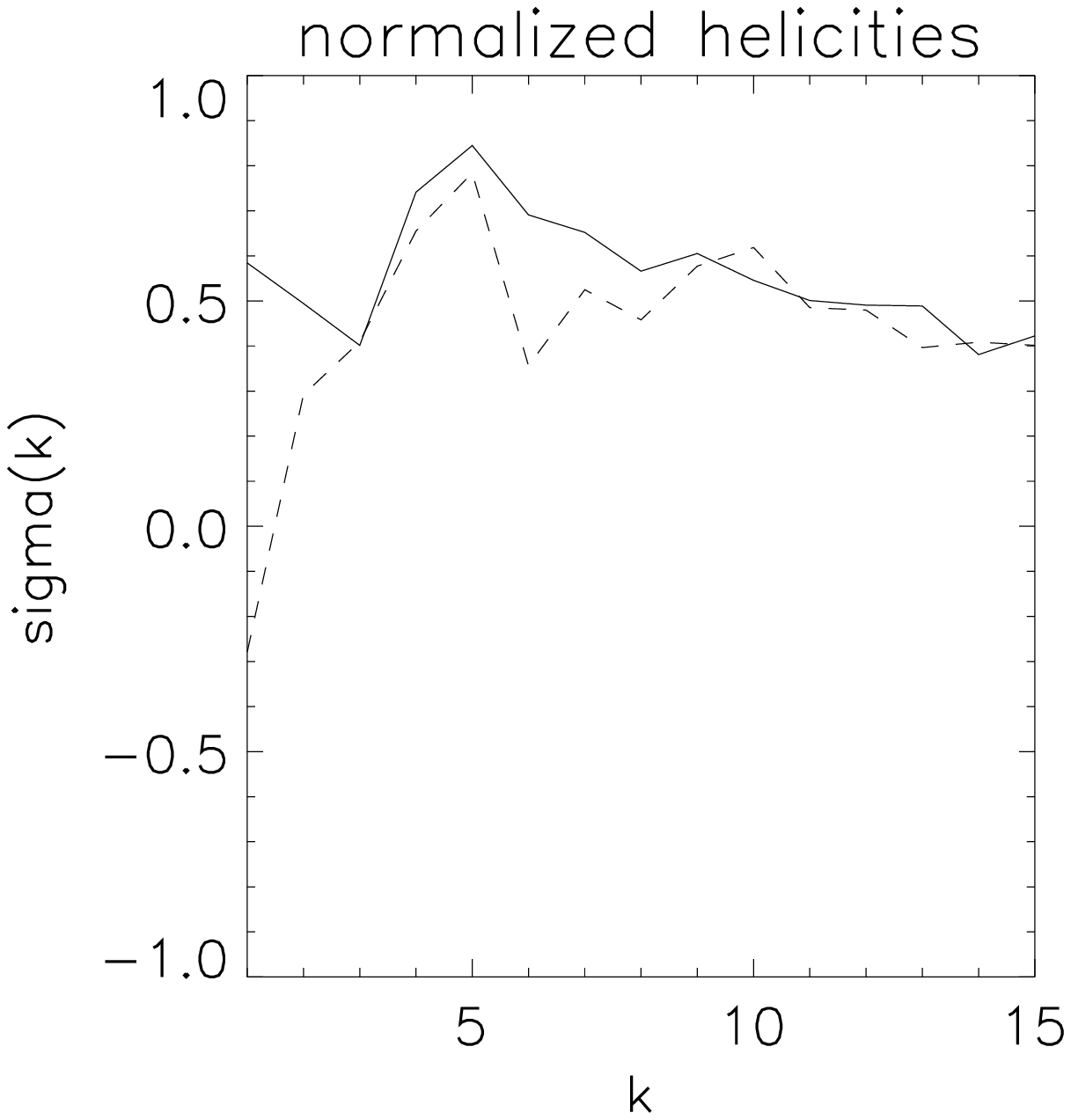]{
 Normalized kinetic and magnetic helicities for the strong dc magnetic field
case at $t=1000$.  There is little helicity of either type at the longest wavelength.
There is no inverse cascade, and dynamo action is suppressed.
\label{fig_6}}
\end{figure}


\begin{references}{} 

\bibitem 
{FrischEA75}
U. Frisch, A. Pouquet, J. Leorat and A. Mazure, 
J. Fluid Mech. {\bf 68}, 769 (1975). 

\bibitem 
{PouquetEA76}
A. Pouquet, U. Frisch and J. Leorat, 
J. Fluid Mech. {\bf 77}, 321 (1976).

\bibitem 
{Pouquet87}
A. Pouquet in {\it Proc. Les Houches Summer School on Astrophysical
 Fluid Dynamics}, ed. by J. P. Jahn and J. Zinn-Justin
 (Elsevier, Amsterdam, 1987) , pp. 139-227.

\bibitem 
{MeneguzziEA81}
M. Meneguzzi, U. Frisch and A. Pouquet, 
Phys. Rev. Lett. {\bf 47}, 1060 (1981).

\bibitem 
{Moffatt78}
H. K. Moffatt, in {\it Magnetic Field Generation in Electrically
Conducting Fluids} (Cambridge: Cambridge University Press, 1978).

\bibitem 
{KrauseRaedler80}
F. Krause and K. H. R\"adler in {\it Mean-Field Magnetohydrodynamics and
 Dynamo Theory} (Oxford: Clarendon Press, 1980).

\bibitem 
{FieldEA99}
G. B. Field, E. G. Blackman and H. Chou, 
Astrophys. J. {\bf 513}, 638 (1999).

\bibitem 
{ShebalinEA83}
J. V. Shebalin, W. H. Matthaeus and D. Montgomery, 
J. Plasma Phys. {\bf 29}, 525 (1983).

\bibitem 
{OughtonEA94}
S. Oughton, E. Priest and W. H. Matthaeus, 
J. Fluid Mech. {\bf 280}, 95 (1994).

\bibitem 
{MontgomeryMatthaeus95}
D. Montgomery and W. H. Matthaeus, 
Astrophys. J. {\bf 447}, 706 (1995).

\bibitem 
{MilanoEA01}
L. J. Milano, W. H. Matthaeus, P. Dmitruk and D. C. Montgomery, 
 Phys. Plasmas {\bf 8}, 2673 (2001).

\bibitem 
{CattaneoHughes96}
F. Cattaneo and D. W. Hughes,
Phys. Rev. E {\bf 54}, 4532 (1996).

\bibitem 
{BlackmanField00}
E. G. Blackman and G. B. Field,
Astrophys. J. {\bf 534}, 984 (2000).

\bibitem 
{BrandenburgDonner97}
A. Brandenburg and K. J. Donner,
Mon. Not. R. Astron. Soc. {\bf 288}, L29 (1997).

\bibitem 
{GottliebOrszag77}
D. Gottlieb and S. A. Orszag in
{\it Numerical Analysis of Spectral Methods: Theory and Applications}
(SIAM, Philadelphia, 1977).

\bibitem 
{Berger97}
M. A. Berger, 
J. Geophys. Res. {\bf 102}, 2637 (1997).

\bibitem 
{MontgomeryChen84}
D. Montgomery and H. Chen, 
Plasma Phys. Controlled Fusion {\bf 26}, 1189 (1984).

\bibitem 
{StriblingMatthaeus94}
T. Stribling, W. H. Matthaeus and S. Ghosh, 
J. Geophys. Res. {\bf 99}, 2567 (1994).

\bibitem 
{MontgomeryBates99}
D. C. Montgomery and J. W. Bates, 
Phys. Plasmas {\bf 6}, 2727 (1999).

\bibitem 
{Matthaeus99}
W. H. Matthaeus, in
{\it Geophysical Mono.\ 111. Proceedings of Magnetic Helicity
in Space and Laboratory Plasmas},
ed. by M. R. Brown, R. C. Canfield and A. A. Pevtsov,
(American Geophysical Union, Washington DC, 1999).

\bibitem 
{ShanEA91}
X. Shan, D. Montgomery and H. Chen, 
Phys. Rev. A {\bf 44}, 6800 (1991).

\bibitem 
{ShanMontgomery93a}
X. Shan and D. Montgomery, 
Plasma Phys. Controlled Fusion {\bf 35}, 619 (1993).

\bibitem 
{ShanMontgomery93b}
X. Shan and D. Montgomery, 
Plasma Phys. Controlled Fusion {\bf 35}, 1019 (1993).

\bibitem 
{ShanMontgomery94}
X. Shan and D. Montgomery, 
Phys. Rev. Lett. {\bf 73}, 1624 (1994).


\end{references}
\end{document}